\documentclass[manuscript]{emulateapj}
\usepackage{amsmath}
\usepackage{amssymb}
\usepackage{natbib}
\usepackage{hyperref}
\usepackage{breakurl}
\usepackage{xcolor}
\usepackage{enumitem}
\setlist{leftmargin=5.5mm}
\usepackage{lineno}

\newcommand{\flux}{\,erg\,cm$^{-2}$\,s$^{-1}$}
\newcommand{\lum}{\,erg\,s$^{-1}$}

\newcommand{\cm}{\,cm$^{-2}$}
\newcommand{\nh}{$N_\mathrm{H}$}

\newcommand{\src}{3FGL~J0954.8$-$3948}
\shorttitle{Multi-wavelength study of 3FGL~J0954.8$-$3948}
\shortauthors{Li et al.}

\begin{document}
\title{Multi-wavelength observations of a new redback millisecond pulsar candidate: 3FGL~J0954.8$-$3948}
\author{
Kwan-Lok Li\altaffilmark{1},
Xian Hou\altaffilmark{2,3},
Jay Strader\altaffilmark{1},
Jumpei Takata\altaffilmark{4},
Albert K. H. Kong\altaffilmark{5},
Laura Chomiuk\altaffilmark{1},
Samuel J. Swihart\altaffilmark{1},
Chung Yue Hui\altaffilmark{6},
and K. S. Cheng\altaffilmark{7}
}
\altaffiltext{1}{Department of Physics and Astronomy, Michigan State University, East Lansing, MI 48824, USA; \href{mailto:liliray@msu.edu}{liliray@msu.edu} (KLL)}
\altaffiltext{2}{Yunnan Observatories, Chinese Academy of Sciences, Kunming, 650216, China}
\altaffiltext{3}{Key laboratory for the Structure and Evolution of Celestial Objects, Chinese Academy of Sciences, Kunming, 650216, China}
\altaffiltext{4}{School of physics, Huazhong University of Science and Technology, Wuhan 430074, China}
\altaffiltext{5}{Institute of Astronomy, National Tsing Hua University, Hsinchu 30013, Taiwan}
\altaffiltext{6}{Department of Astronomy and Space Science, Chungnam National University, Daejeon 34134, Korea}
\altaffiltext{7}{Department of Physics, The University of Hong Kong, Pokfulam Road, Hong Kong}

\begin{abstract}
We present a multi-wavelength study of the unassociated \textit{Fermi}-LAT source, \src, which is likely the $\gamma$-ray counterpart of a 9.3-hour binary in the field. With more than 9 years of Pass 8 LAT data, we updated the $\gamma$-ray spectral properties and the LAT localization of the $\gamma$-ray source. 
While the binary lies outside the cataloged 95\% error ellipse, the optimized LAT ellipse is 0.1 degrees closer and encloses the binary. 
The system is likely spectrally hard in X-rays (photon index $\Gamma_X=1.4^{+1.2}_{-1.0}$) with orbital modulations detected in optical, UV, and possibly X-rays. A steep spectrum radio counterpart (spectral index $\alpha\approx-1.6$) is also found in the TIFR GMRT Sky Survey (TGSS), implying that it is a pulsar system. We obtained a series of SOAR and Gemini spectroscopic observations in 2017/2018, which show a low-mass secondary orbiting in a close circular orbit with $K_2=272\pm4$~km s$^{-1}$ under strong irradiation by the primary compact object. All the observations as well as the modelling of the X/$\gamma$-ray high-energy emission suggest that \src\ is a redback millisecond pulsar in a rotation-powered state. 
\\ 
\end{abstract}

\section{Introduction}
Redback and black widow millisecond pulsars (MSPs) are unique subclasses of pulsar binaries, which have compact orbits (periods of $\lesssim$ 1 days) and very low-mass companions ($M_2=$ 0.1 -- 0.4 $M_\odot$ for redback and $<0.1M_\odot$ for black widow; \citealt{2013IAUS..291..127R,2013ApJ...775...27C}). With separations of just a few $R_\odot$ or less, the primary pulsars are heavily ablating the secondary stars with high-power radiation and pulsar winds, potentially explaining how isolated MSPs are formed \citep{1988Natur.334..227V}. They became even more interesting recently as three redback MSPs (also known as transitional MSPs), M28I \citep{2013Natur.501..517P}, PSR J1023+0038 \citep{2009Sci...324.1411A,2014ApJ...781L...3P,2014ApJ...790...39S}, and PSR J1227$-$4853 \citep{2015ApJ...800L..12R}, have shown remarkable transitions between the low-mass X-ray binary (LMXB) state and the radio pulsar state, which could be an important piece of evidence for the recycling explanation for MSP formation \citep{1982Natur.300..728A}. 

In the \textit{Fermi} Large Area Telescope (LAT) third source catalog (3FGL; \citealt{2015ApJS..218...23A}), over 1000 $\gamma$-ray sources are without definitive associations at other wavelengths. 
On the basis of previous work (e.g., \citealt{2013ApJS..208...17A}), some of these unassociated Fermi-LAT sources are likely to be redback and black widow MSPs. 
In fact, recent coordinated multi-wavelength searches have identified at least nine unassociated \textit{Fermi}-LAT sources as promising redback/black widow candidates, including\footnote{The list does not include 2FGL~J0846.0+2820 \citep{2017ApJ...851...31S} and 1FGL~J1417.7$-$4407 \citep{2015ApJ...804L..12S}, which have giant secondaries with orbital periods of 8.1 days and 5.4 days, respectively. 
The latter was confirmed as a MSP (PSR J1417$-$4402; \citealt{2016ApJ...820....6C})
}: 
3FGL J0838.8$-$2829 \citep{2017ApJ...844..150H},
3FGL J0212.1+5320 \citep{2016ApJ...833..143L,2017MNRAS.465.4602L},
3FGL J0427.9$-$6704 \citep{2016ApJ...831...89S},
3FGL J2039.6$-$5618 \citep{2015ApJ...812L..24R,2015ApJ...814...88S},
3FGL J1544.6$-$1125 \citep{2015ApJ...803L..27B},
2FGL J1653.6$-$0159 \citep{2014ApJ...793L..20R,2014ApJ...794L..22K},
1FGL J0523.5$-$2529 \citep{2014ApJ...788L..27S},
2FGL J1311.7$-$3429 \citep{2012ApJ...754L..25R,2012ApJ...757..176K},
1FGL J2339.7$-$0531 \citep{2011ApJ...743L..26R,2012ApJ...747L...3K},
at least two of which, PSR J1311-3430 \citep{2012Sci...338.1314P} and PSR J2339-0533 \citep{2015ApJ...807...18P}, have been confirmed as MSPs. It is also worth mentioning that 3FGL J0427.9$-$6704 and 3FGL J1544.6$-$1125 are $\gamma$-ray emitting low-mass X-ray binaries that are good candidates for transitional MSPs.

\src\ is a bright unassociated \textit{Fermi}-LAT source in 3FGL, detected in 0.1--100~GeV with a detection significance of $19\sigma$ \citep{2015ApJS..218...23A}. It first appeared in the \textit{Fermi}-LAT first source catalog (1FGL; \citealt{2010ApJS..188..405A}), and subsequently in the \textit{Fermi}-LAT second source catalog (2FGL; \citealt{2012ApJS..199...31N}). The $\gamma$-ray source resembles other $\gamma$-ray pulsars with low source variability (its chi-squared variability index of 51 with 47 degrees of freedom is consistent with a steady source)
and a significantly curved $\gamma$-ray spectrum \citep{2013ApJS..208...17A}. Both features suggest that \src\ could be a $\gamma$-ray pulsar. Indeed, \cite{2016ApJ...820....8S} found that \src\ is a strong MSP candidate using statistical and machine learning techniques. 
\cite{2017MNRAS.469.3688D} suggested a binary with an orbital period of 9.3 hours, named SSS~J095527.8$-$394752, to be a promising counterpart to 1FGL~J0955.2$-$3949. In addition, a bright radio counterpart possibly associated with the binary was independently discovered by \cite{2016MNRAS.461.1062F}. 
However, these counterparts are later found to be outside the 3FGL error ellipse, and the association therefore remains questionable until now. 

In this paper, we present multi-wavelength observations of \src, with which we show that SSS~J095527.8$-$394752 is still likely to be the counterpart of the $\gamma$-ray source. Furthermore, the observed timing and spectral properties are in agreement with the suggestion of \cite{2017MNRAS.469.3688D} that \src\ is a new member of the group of redback MSPs. 

\section{\textit{Fermi}-LAT gamma-ray analysis}

\setlength{\tabcolsep}{3pt}
\begin{table*}
\centering 
\caption{\textit{Fermi}-LAT Properties of \src}
\begin{tabular}{l|cc|ccccc|cccc}
\toprule %
Model & \multicolumn{2}{c|}{3FGL Position} & \multicolumn{5}{c|}{Re-localization (1--100~GeV)} & \multicolumn{4}{c}{Best-fit Parameters (0.1--100~GeV)} \\
\hline
& $-\log \mathcal{L}$ & TS & R.A.$\rm^a$ & Dec.$\rm^a$ & 95\% radius & $-\log \mathcal{L}$ & TS & $\Gamma_g$ & $E_c$ & $\beta$ & $F_{\rm ph}$\\
& & & (degree) & (degree) & (degree) & & & & (GeV) & & ($10^{-8}$ cm$^{-2}$ s$^{-1}$)\\
\hline
PLExpCutoff & 125311.76 & 309 & $148.8462$ & $-39.8089$ &$0.0369$ & 125289.08 &355 & $2.2\pm0.1$ & $4.5\pm1.2$ & \nodata & $3.1\pm0.2$\\
LogParabola & 125312.25 & 309 & $148.8459$ &$-39.8087$ &$0.0368$ & 125289.84 &353 & $2.2\pm0.1$ & \nodata & $0.13\pm0.04$ & $2.9\pm0.2$\\
\hline
\multicolumn{12}{l}{$\rm^a$The coordinates are in the J2000 frame. }\\
\end{tabular}
\label{fermi_fit}
\end{table*}

\begin{figure*}
\centering
\includegraphics[width=1.0\textwidth]{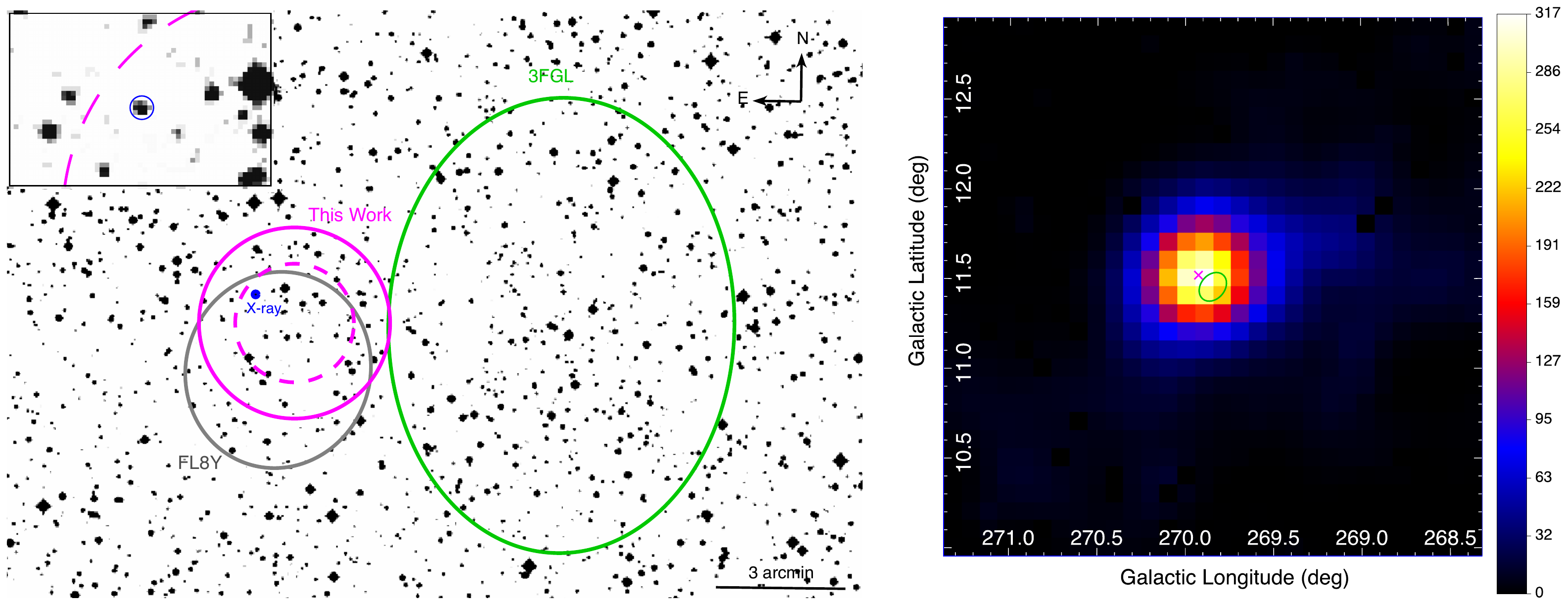}
\caption{Left: DSS-1 image of the field of \src\ overlapped with: green and gray ellipses as the 95\% error regions of \src\ in 3FGL and FL8Y, respectively; magenta concentric circles as the new LAT error circles in 68\% (dashed) and 95\% (solid) confidence levels (PLExpCutoff spectral model assumed); small blue circle as the 90\% error circle of 1SXPS~J095527.8$-$394750. The upper left inset box is the zoomed-in view of the optical counterpart, SSS~J095527.8$-$394752. Right: TS map of \src\ (in Galactic coordinates) with the 3FGL 95\% error ellipse (green) and the new LAT best-fit position (magenta cross). The new position is consistent with the brightest central pixel, which signifies the highest source detection significance. 
}
\label{fig:dss}
\end{figure*}

Before searching for the X-ray/optical counterparts, we first re-localized \src\ in $\gamma$-rays taking advantage of the Pass 8 LAT data \citep{2013arXiv1303.3514A} with a timespan of more than 9 years, whose performance has been much improved over the 4 years of Pass 7 data used in 3FGL. This would in principle yield a more precise $\gamma$-ray position with a more restricted error circle. 

The LAT events and the spacecraft data were downloaded from the {\it Fermi} Science Support Center (FSSC).\footnote{\url{http://fermi.gsfc.nasa.gov/ssc/}}
The dataset covers the time range of 4 Aug 2008 -- 16 Dec 2017 with reconstructed energy in 0.1--100 GeV. We selected SOURCE class events (Front and Back) but excluded those with a zenith angle larger than $90^\circ$ or a rocking angle larger than $52^\circ$ to avoid the Earth limb contamination. The region of interest (ROI) was centered at $(\alpha,\delta)=(148\fdg712,-39\fdg809)$, which is the 3FGL position of \src. All the 3FGL sources located within a $20^\circ$ radius circle at the center were included to build a spatial and spectral model for the $\gamma$-ray emission in the field. The model also includes the latest Galactic interstellar emission model (gll\_iem\_v06.fits) and the isotropic emission spectrum (iso\_P8R2\_SOURCE\_V6\_v06.txt), of which the latter takes the extragalactic emission and the residual instrumental background into account \citep{2016ApJS..223...26A}. 

With the latest instrument response function P8R2\_SOURCE\_V6, model fitting was performed by the maximum likelihood method \citep{mattox96}, which is integrated in the {\it Fermi} Science Tools available at FSSC.
In addition, the $\mathtt{fermipy}$ package\footnote{\url{http://fermipy.readthedocs.io/en/latest/index.html}} developed within the LAT collaboration was used to facilitate the analysis \citep{2017arXiv170709551W}. 
Under the framework, the significance of a certain source is characterized by Test Statistic (TS), 
\begin{equation}\nonumber
{\rm TS} =2(\log \mathcal{L}-\log \mathcal{L}_{0}), 
\end{equation}
where $\log \mathcal{L}$ and $\log \mathcal{L}_{0}$ are the logarithms of the maximum likelihood of the complete source model and of the null hypothesis model (i.e., the source model without the certain source), respectively. 

The localization fit was performed at $>1$ GeV to benefit from the improved angular resolution and the reduced background contamination. 
We first performed an initial fit with a $14^\circ \times 14^\circ$ ROI, by allowing the background diffuse components as well as all the sources located $<5^\circ$ from \src\ to vary. 
Given the $\sim4\sigma$ evidence for spectral curvature of \src\ in 3FGL, we employed two curved spectral models for it, 
\begin{equation}\nonumber
\begin{split}
&{\rm PLExpCutoff:}\qquad\dfrac{dN}{dE} \propto E^{-\Gamma_g} \exp\Big(-\frac{E}{E_c}\Big)\\
{\rm and}\\
&{\rm LogParabola:}\qquad\dfrac{dN}{dE} \propto \Big(\frac{E}{E_b}\Big)^{-(\Gamma_g+\beta\log(E/E_b))},
\end{split} 
\end{equation}
where $\Gamma_g$ serves as the photon index for both models, $E_c$ characterizes the cutoff energy for PLExpCutoff, $\beta$ defines the degree of curvature for LogParabola, and $E_b$ is a fixed scale parameter. 
Both models give good fits to the data and PLExpCutoff is slightly preferred over LogParabola by $1\sigma$. 
Three parameter-fixed $\gamma$-ray sources located outside the $5^\circ$ region, 3FGL~J0928.9$-$3530, 3FGL~J0937.1$-$4544c and 3FGL~J1007.4$-$3334 were not well modelled in the first round and a second iteration was done with their normalization parameters freed. 

Taking the best-fit source model from the aforementioned processes as the input, we re-localized \src\ using the ``Source localization" function in $\mathtt{fermipy}$ for both LogParabola and PLExpCutoff models. The spectral parameters of all the sources, except the Galactic/Isotropic diffuse components and \src, were fixed during the localization, although we found that freeing the normalizations of the background diffuse components has no significant effect on the localization result. 
Table~\ref{fermi_fit} summarizes the best localization parameters of \src\ for PLExpCutoff and LogParabola.
There is no obvious difference between the localizations from LogParabola and PLExpCutoff, but the PLExpCutoff version is slightly favoured given the smaller $-\log \mathcal{L}$ of the best-fit. 
Figure~\ref{fig:dss} shows the new $\gamma$-ray localization of \src\ (PLExpCutoff) on the DSS1 image and the TS map of the field. 
Compared with the 95\% error ellipse of \src\ presented in 3FGL (i.e., 5\farcm3 and 4\farcm0 for the semi-major and semi-minor axes, respectively), the updated LAT position is shifted by 6\farcm2 to $(\alpha,\delta)=(148\fdg8462,-39\fdg8089)$ with a much improved 95\% error radius of 2\farcm2 (Figure \ref{fig:dss}). 
Using this best-fit position, broadband spectral fitting in the energy range of 0.1--100~GeV was also performed for both PLExpCutoff and LogParabola, and the best-fit parameters are listed in Table \ref{fermi_fit}. 

\begin{figure}
\centering
\includegraphics[width=85mm]{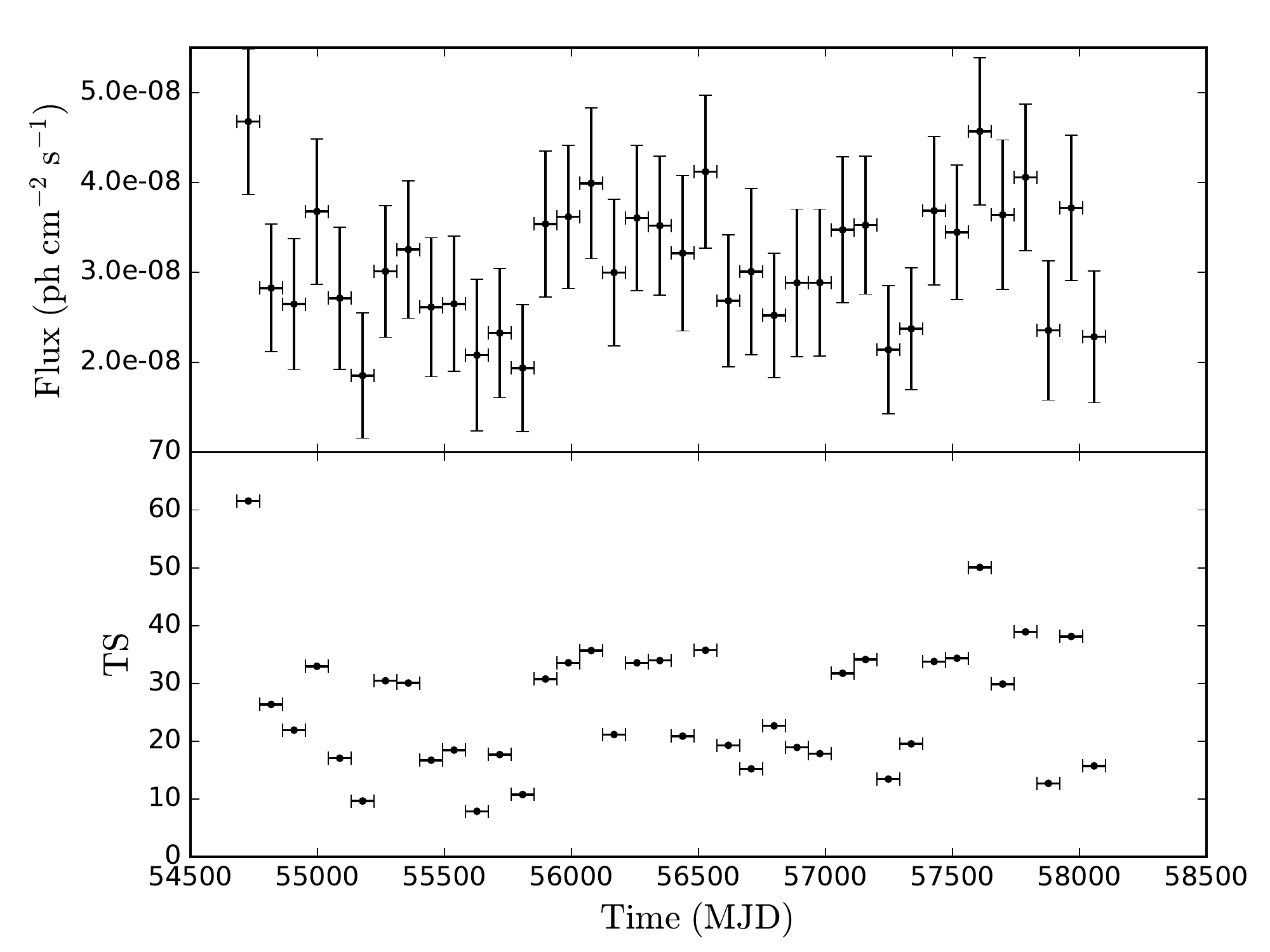}
\caption{Light curve and TS evolution for \src\ in the energy range of $0.1-100$ GeV with 90-day binning.}
\label{fermi_lc}
\end{figure}

High long-term $\gamma$-ray flux variability is a common feature of transitional MSPs. To examine this possibility for \src, we computed a long-term light curve with 90-day binning in the energy range of $0.1-100$ GeV (Figure \ref{fermi_lc}). For each time bin, the flux was calculated independently using the likelihood analysis. In the source model used, the spectral shapes of all sources (including \src) located within $5^\circ$ from \src\ were fixed according to their best-fit models determined previously (i.e., only normalizations were allowed to vary). Upper limits at 95\% confidence level were calculated when \src\ had TS $<4$. In addition, we followed the method presented in \cite{2015ApJS..218...23A} to obtain the variability significance of \src, which is just $0.4\sigma$ (for 37 degrees of freedom). The $\gamma$-ray emission is thus considered to be stable on a time scale of a few months, ruling out the possibility of \src\ being a transitional MSP within the LAT mission. 



\section{\textit{Neil Gehrels Swift} Observations}
\label{sec:swift}
\textit{Swift} has observed the field of \src\ 10 times between March 2010 and March 2017. Five of the observations taken in 2015 are with short exposure times less than 400~sec while the other five have relatively long exposures (Table~\ref{tab:swift}). In the stacked XRT image, only one X-ray source at $\alpha\mathrm{(J2000)}=09^\mathrm{h}55^\mathrm{m}28\fs40$, $\delta\mathrm{(J2000)}=-39\arcdeg 48\arcmin 02\farcs1$ (90\% positional uncertainty: 3\farcs5) is detected within the new 95\% LAT error circle (Figure \ref{fig:dss}).
The X-ray source is listed as 1SXPS~J095527.8$-$394750 in the \textit{Swift}-XRT point source catalog (1SXPS; \citealt{2014ApJS..210....8E}), in which only part of the observations was analysed. We therefore re-analysed the source with all ten observations to improve the spectral fitting as well as the long-term X-ray light curve of the source. As the field is not crowded and no other bright X-ray source is found around the target, we simply used the \textit{Swift}/XRT on-line tools\footnote{\url{http://www.swift.ac.uk/user_objects/}} provided by the \textit{Swift} team \citep{2007A&A...469..379E,2009MNRAS.397.1177E} to perform the following XRT analyses. 

While 1SXPS~J095527.8$-$394750 is significantly detected, the low photon statistics only allow a basic spectral analysis (i.e., only 32 photons collected within a 20\arcsec\ radius circular region). To deal with the low-count spectrum, we used \texttt{XSPEC} (version 12.9.1m) with \textit{W-statistic} (a modified version of \textit{C-statistic}; \citealt{1979ApJ...228..939C}) in the fitting process. 
In addition, we binned the spectrum accordingly so that every bin contains at least one source count as the development team suggested. 
A simple absorbed power-law is assumed and the best-fit parameters are \nh\ $=5.2^{+8.9}_{-5.0}\times10^{21}$\cm, $\Gamma_X=1.4^{+1.2}_{-1.0}$, and $F_{\rm 0.3-10keV}=3.0^{+3.6}_{-1.0}\times10^{-13}$\flux\ (absorption corrected; all the uncertainties listed are in 90\% confidence interval). 
Although it is not significant, the best-fit \nh\ is a few times higher than the Galactic value of $1.34\times10^{21}$\cm\ \citep{2005A&A...440..775K}, may indicating an intrinsic absorption of the system. If a value closer to the Galactic value is used, e.g., \nh\ $=10^{21}$\cm, the best-fit parameters change to $\Gamma_X=0.8\pm0.5$ and $F_{\rm 0.3-10keV}=3.1^{+1.9}_{-1.2}\times10^{-13}$\flux. 
We also tried the thermal model \texttt{mekal}, which requires an unreasonably high temperature of $kT\approx30$~keV to describe the spectrum. Given the poor data quality of the spectrum, we did not try any complex models with two or multiple emission components. 


\begin{table*}
\centering 
\caption{\small \textit{Swift} Observations of \src}
\begin{tabular}{lllrrcr}
\toprule
\textit{Swift} ObsID & Date & BJD & XRT Exposure & XRT Count Rate$^{\rm a}$ & UVOT Filter & Magnitude$^{\rm a}$\\
 & & (start time) & (seconds) & ($10^{-3}$ cts s$^{-1}$) & & (Vega) \\
\hline
00031664001 & 2010 Mar 24 & 2455279.5230275 & 3557 & $5.0\pm1.4$ & UVW1 & $20.24\pm0.13$\\
00084699001 & 2015 Feb 04 & 2457057.7715524 & 238 & $<27.0^{\rm b}$ & UVW2 & $>20.40$ \\
00084699002 & 2015 Feb 16 & 2457070.4008172 & 145 & $<37.0^{\rm b}$ & UVW2 & $>20.33$ \\
00084699003 & 2015 Jun 21 & 2457195.0878701 & 186 & $<24.7^{\rm b}$ & UVM2 & $>20.78$ \\
00084699004 & 2015 Aug 07 & 2457242.3099548 & 278 & $<22.5^{\rm b}$ & UVW2 & $>20.47$ \\
00084699005 & 2015 Aug 12 & 2457247.3567788 & 356 & $<36.1^{\rm b}$ & UVM2 & $>20.36$ \\
00084699006 & 2015 Aug 19 & 2457254.1462335 & 1578 & $<8.7^{\rm b}$ & UVW2 & $20.07\pm0.19$ \\
00084699007 & 2015 Aug 24 & 2457258.9278091 & 1091 & $4.4^{+2.8}_{-2.0}$ & UVM2 & $20.01\pm0.24$ \\
00034854001 & 2016 Dec 22 & 2457744.5423559 & 1785 & $3.4^{+2.0}_{-1.5}$ & U & $18.97\pm0.07$\\
00034854002 & 2017 Mar 09 & 2457821.8875329 & 2008 & $3.9^{+2.1}_{-1.6}$ & UVW2 & $>21.74$ \\
\hline
\multicolumn{6}{l}{$^{\rm a}$The upper limits listed are in 95\% confidence level.} \\
\multicolumn{6}{l}{$^{\rm b}$The source is detected in the stacked \textit{Swift}/XRT image with $2.1^{+1.6}_{-1.2}\times 10^{-3}$ cts s$^{-1}$.} \\
\end{tabular}
\label{tab:swift}
\end{table*}

Five XRT observations got only low exposure (i.e., much less than 1~ksec) and the X-ray source was therefore undetected in these datasets. Surprisingly, we also found that the source was undetected in a ``deep'' observation taken on 2015 August 19 with an exposure time of about 1.6~ksec (Table~\ref{tab:swift}). 
Within a 47\arcsec\ radius circular region centered at the source position (corresponding to 90\% of the encircled energy fraction of XRT at 1.5 keV; \citealt{2005SPIE.5898..360M}), only one photon (which is located near the edge of the region) was detected in this 1.6~ksec observation. Even assuming that this only event is from the source, the inferred count rate is much lower than the measurements in 2010 and 2015--2017---for example, seven source counts would have been detected in a 1.6~ksec observation with the count rate of $4.4\times10^{-3}$~cts~s$^{-1}$ measured five days later (Table~\ref{tab:swift}). Using a Bayesian approach \citep{1991ApJ...374..344K}, we computed 95\% upper limits for all the non-detections. As expected, the upper limits for data with $<1$~ksec are not very much constraining (i.e., a few $\times10^{-2}$~cts~s$^{-1}$, while the average count rate of the four individual detections is about $4\times10^{-3}$~cts~s$^{-1}$). The upper limit for the 1.6~ksec data is deeper (i.e., $<8.7\times10^{-3}$~cts~s$^{-1}$), however, still insufficient to clarify whether the low-count-rate measurement is physically- or statistically-based. For a deeper constraint, we combined all the six XRT observations and the X-ray source can be marginally detected in the stacked image with $2.1^{+1.6}_{-1.2}\times 10^{-3}$ cts s$^{-1}$. Although this marginal detection shows a $\sim50$\% decrease on flux in the period from 2015 February 04 through August 19, the variability is not statistically significant (i.e., less than 2$\sigma$). To check whether this variability was seen at other frequencies, we performed a \textit{Fermi}-LAT analysis with the data collected between 2015 February 04 and August 19, and the $\gamma$-ray flux (100~MeV--100~GeV) did not vary significantly. 
In UV, there are some UVOT images taken simultaneously with the XRT observations. Although the UVOT magnitudes (obtained by aperture photometry using the \texttt{uvotsource} task in \texttt{HEAsoft} v6.22) were significantly changing over time (Table~\ref{tab:swift}), it is due to the orbital modulation (Figure~\ref{fig_fold_lc}; will be discussed in the coming sections). 

\section{Catalina Surveys Data}
\label{sec:crts}
\begin{figure}
\centering
\includegraphics[width=0.48\textwidth]{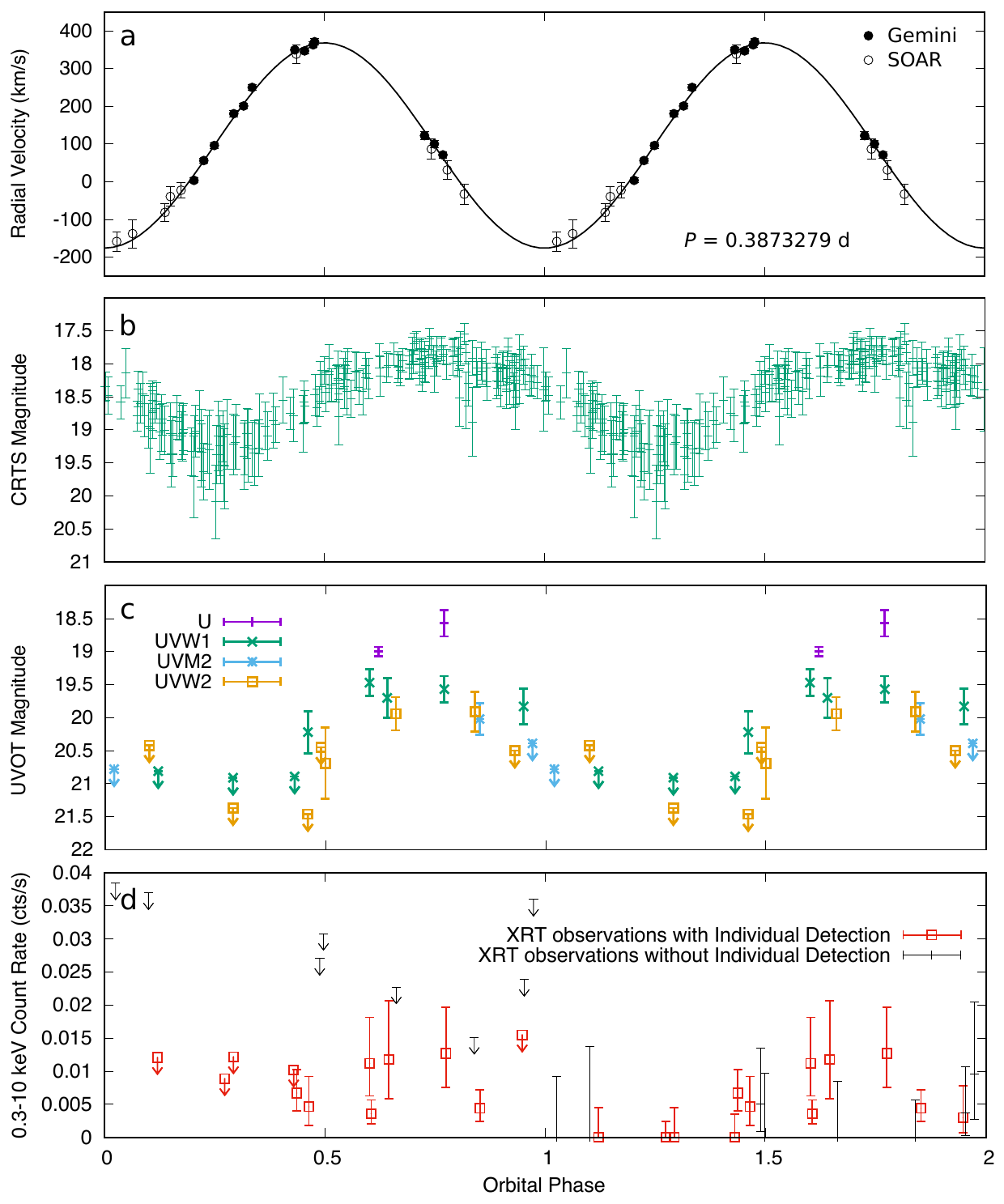}
\caption{The figure shows (a) the radial velocities curve, (b)
the CSS light curve, (c) the UVOT light curves, and (d) the \textit{Swift}/XRT (0.3--10~keV) light curve (binned per orbit), of \src, which are all folded on the orbital period of $P_{\rm joint}=0.3873279$ d with phase zero at BJD 2458097.94395 (the ascending node). 
Two cycles are shown for clarity. 
In the panel (d), 95\% upper limits and net count rates (even zero net count rate) are shown in the first and second cycles, respectively, for bins with insignificant detection. 
}
\label{fig_fold_lc}
\end{figure}

In the CSS Periodic Variable Star Catalogue, \cite{2017MNRAS.469.3688D} identified a $V_{\rm css}=18.45$~mag candidate optical counterpart for the X-ray system, SSS~J095527.8$-$394752 (the accurate \textit{Gaia} position in the Data Release 2: $\alpha\mathrm{(J2000)}=09^\mathrm{h}55^\mathrm{m}27\fs8090842\pm0.11\,$mas, $\delta\mathrm{(J2000)}=-39\arcdeg 47\arcmin 52\farcs29613\pm0.13\,$mas; \citealt{2016A&A...595A...1G,2018arXiv180409365G}), which is the only optical source located within the 90\% positional uncertainty of 1SXPS~J095527.8$-$394750. The source was classified as a non-EA (Algol type) eclipsing binary with $P_{\rm css}=0.387330$ d (about 9.3 hours). The photometric data (186 individual exposures) obtained from the Catalina Surveys
Data Release 2 (CSDR2; \citealt{2009ApJ...696..870D})\footnote{All the CSS data was taken unfiltered.} shows a clear sinusoidal-like profile at 9.3 hours (Figure~\ref{fig_fold_lc}; the light curve has been barycentric corrected to the TDB system; \citealt{2010PASP..122..935E}). We fit the data with a sinusoidal function and the best-fit mean magnitude is $V_{\rm css}\approx18.4$~mag with a peak-to-peak amplitude of $\Delta V_{\rm css}\approx1.2$~mag. For the XRT variability discussed in \S\ref{sec:swift}, no further investigation is allowed on the CSS optical light curve as it does not cover the time of interest.

\section{SOAR and Gemini South Spectroscopy}
\label{sec:soar}
We obtained optical spectroscopy of the source using the Goodman Spectrograph \citep{2004SPIE.5492..331C} on the SOAR telescope (UT 2017 Jul 11 to 2018 Jan 22) and using GMOS-S on the Gemini South telescope (2017 Dec 8 to 19). The SOAR spectra all used a 400 l mm$^{-1}$ grating with a 0\farcs95 slit, giving a FWHM resolution of about 5.3 \AA. Most of the SOAR spectra covered a wavelength range of about 4850 to 8850 \AA, but the last two had redder coverage, from 5950 to 9950 \AA. These nine individual spectra all had exposure times of 1200 sec. The Gemini spectra used the R400 grating, centered around 6800 \AA, with a 1\farcs0 slit. All the exposures were 600 sec. Fifteen spectra were obtained with Gemini; two of these Gemini spectra had signal-to-noise too low to be useful, but the remaining thirteen spectra were good. Both the SOAR and Gemini spectra were reduced and optimally extracted in the usual manner.

The immediate impression upon viewing the spectra is that the effective temperature of the star varies substantially over its orbit, and is quite warm when the ``day" side of the star is facing Earth, suggesting the presence of heating. Comparing to the Paschen series of stars from the Ca triplet library of \cite{2001MNRAS.326..959C}, the Gemini spectra around $\phi \sim 0.75$ are of mid-A type, with effective temperatures around $\sim 8000$ K. The Paschen series entirely disappears by the ``night" phase of $\phi \sim 0.25$, where the relative strength of the metal to hydrogen lines are consistent with an early to mid G-type spectrum, with estimated temperatures of $\sim 5700$ K. The spectra show no evidence of emission lines at any phase.

\begin{figure}
\centering
\includegraphics[width=0.48\textwidth]{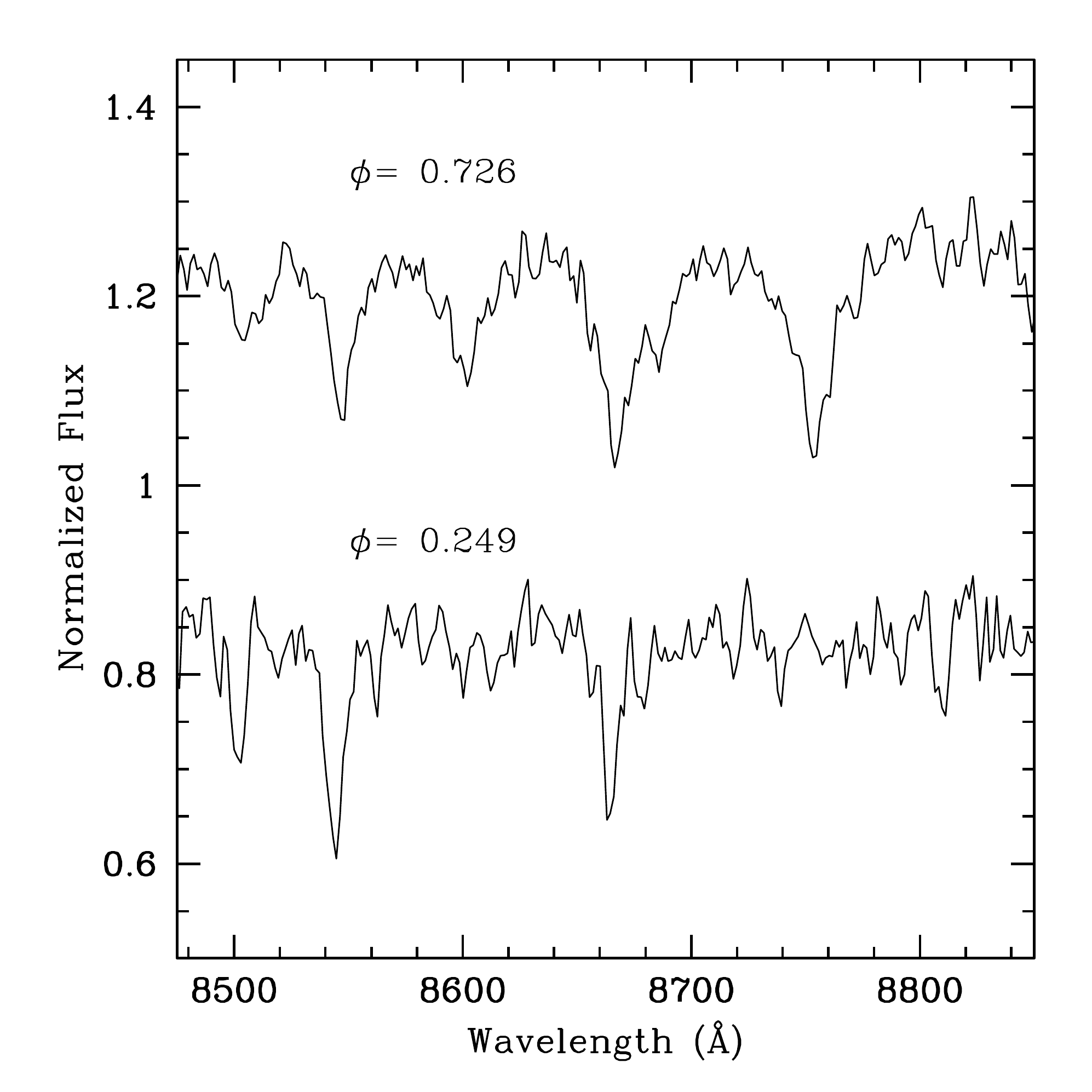}
\caption{Gemini spectra of the star in the red near the extremes of the ``day'' ($\phi = 0.73$) and ``night'' ($\phi = 0.25$) side spectra, with the Paschen series evident on the day side and the Ca triplet on the night side.
}
\label{fig:opt_spec}
\end{figure}

We determined barycentric radial velocities by cross-correlating the individual spectra with templates appropriate to their specific spectral types, primarily in the region of the Ca triplet and the Paschen series. These 22 velocities are listed as barycentric dynamical times on the TDB system \citep{2010PASP..122..935E} in Table \ref{tab:bigg}.

We fit a circular Keplerian model to these radial velocities using the custom Markov Chain Monte Carlo sampler \emph{The Joker} \citep{2017ApJ...837...20P}. The posterior distributions are approximately normal and uncorrelated. We summarize these with medians and equivalent $1\sigma$ quantiles: period $P_{\rm spec} = 0.3873396(81)$ d, semi-amplitude $K_2 = 272\pm4$ km s$^{-1}$, systemic velocity $\gamma = 96\pm3$ km s$^{-1}$, and time of ascending node for the compact object 
$T_{\rm 0,spec} = 2458097.94433\pm0.00053$ d. 
A fit with these median values has $\chi^2$/d.o.f. = 17.7/18, suggesting a reasonable fit. The orbital period is consistent with the CSS value (\S \ref{sec:crts}). Dropping the assumption of a circular orbit does not significantly improve the fit.

Assuming that the orbital period derivative over the period between the CSS and the SOAR/Gemini epochs is negligible (i.e., a stable period), we further constrain the orbital period by performing a joint fitting of the CSS photometric data and the radial velocities with a sinusoidal function and a cosinusoidal function (with the same period and $T_0$ parameters shared), respectively. The best-fit orbital period is $P_{\rm joint} = 0.3873279(3)$ d with $T_{\rm 0,joint} = 2458097.94395\pm0.00068$ d ($\chi^2$/d.o.f. = 134.9/202). This more precise orbital solution was used in the UVOT and XRT timing analyses discussed in \S\ref{sec:swift} and \S\ref{sec:xmod}, respectively (Figure~\ref{fig_fold_lc}). In addition, we tried fitting the data by allowing a ``phase shift'' between the CSS photometric light curve and the radial velocity curve, which is possible when the heating pattern is asymmetric. This gives $P_{\rm joint,s} = 0.3873318(13)$ d with $T_{\rm 0,joint,s} = 2458097.94384\pm0.00068$ d and $\Delta\phi=0.08\pm0.03$ (i.e., the photometric peak at $\phi=0.83$; $\chi^2$/d.o.f. = 125.6/201). 

\begin{table}
\centering 
\caption{Barycentric Radial Velocities of \src \label{tab:bigg}}
\begin{tabular}{crrr}
\toprule
BJD & Radial Vel. & Unc. & Source \\
(d) & (km s$^{-1}$) & (km s$^{-1}$) & \\
\hline
2457945.5057483 & 338.1 & 20.2 & SOAR \\
2457956.4698649 & 86.6 & 21.3 & SOAR \\
2457956.4838969 & 31.4 & 20.1 & SOAR \\
2457956.4988242 & -32.9 & 22.1 & SOAR \\
2458072.7785401 & -158.3 & 21.7 & SOAR \\
2458072.7925358 & -137.4 & 34.5 & SOAR \\
2458072.8257413 & -39.0 & 20.2 & SOAR \\
2458095.7339401 & 180.2 & 8.5 & Gemini \\
2458095.7425900 & 200.8 & 8.2 & Gemini \\
2458095.7501254 & 249.8 & 7.6 & Gemini \\
2458097.7417328 & 371.2 & 7.8 & Gemini \\
2458102.7595228 & 349.6 & 7.1 & Gemini \\
2458102.7681525 & 346.9 & 7.1 & Gemini \\
2458102.7756881 & 362.7 & 7.4 & Gemini \\
2458103.8328295 & 3.6 & 6.7 & Gemini \\
2458103.8414655 & 56.2 & 6.8 & Gemini \\
2458103.8507936 & 96.0 & 7.0 & Gemini \\
2458106.7470583 & 122.3 & 9.8 & Gemini \\
2458106.7557001 & 100.0 & 9.7 & Gemini \\
2458106.7632348 & 71.0 & 8.9 & Gemini \\
2458140.6035185 & -81.1 & 23.9 & SOAR \\
2458140.6175631 & -22.2 & 20.1 & SOAR \\
\hline
\end{tabular}
\end{table}


\section{Radio Measurements in the Literature}
Using the Giant Metrewave Radio Telescope (GMRT) 150~MHz All-Sky Radio Survey (TGSS ADR; \citealt{2017A&A...598A..78I}), \cite{2016MNRAS.461.1062F,2018MNRAS.475..942F} identified radio pulsar candidates associated with unidentified Fermi-LAT sources in 2FGL, 3FGL, and the preliminary 8-year source catalog\footnote{\url{https://fermi.gsfc.nasa.gov/ssc/data/access/lat/fl8y/}} (FL8Y). Three candidates were found within 2FGL sources, and one of these is associated with \src. 
However, as the radio source is outside the 3FGL source's 95\% error ellipse, no detailed investigation was done.

The \src-associated pulsar candidate is located at $\alpha\mathrm{(J2000)}=09^\mathrm{h}55^\mathrm{m}27\fs75\pm2\farcs0$, $\delta\mathrm{(J2000)}=-39\arcdeg 47\arcmin 51\farcs1\pm2\farcs7$ (1$\sigma$ uncertainty; ${\rm 2DRMS}=3\farcs4$), which is roughly consistent with the \textit{Gaia} optical position of SSS~J095527.8$-$394752 (the offset: 2\farcs9). 
This radio counterpart is a bright point source with 77~mJy at 150~MHz, but was cataloged neither in the Sydney University Molonglo Sky Survey (SUMSS at 843~MHz; \citealt{2003MNRAS.342.1117M}) nor the NRAO VLA Sky Survey (NVSS at 1.4~GHz; \citealt{1998AJ....115.1693C}), indicating that it has a steep radio spectrum. \cite{2016MNRAS.461.1062F} estimated the spectral index to be $\alpha\approx-1.6$ (where the flux density $S_\nu\propto\nu^\alpha$), although it is highly uncertain given the poor flux constraints by SUMSS and NVSS. Nevertheless, this spectral index is fully consistent with that of a typical pulsar: $\alpha=-1.60\pm0.03$, the weighted mean spectral index of the 441 pulsars recently studied in \cite{2018MNRAS.473.4436J}, for instance. 

\src\ is also one of the 56 targets in the Parkes radio MSP survey \citep{2015ApJ...810...85C}. Unfortunately, no radio pulsation was found in a blind search with five 60-minute observations at 1.4~GHz. However, the offset between the optimized position of the radio search and the radio counterpart is as large as 5\arcmin, which is comparable to the Parkes beam of $7\farcm2$ (half-width at half-maximum; \citealt{2015ApJ...810...85C}). 
This offset could have reduced the sensitivity of the search. 
Additionally, the radio pulsations could be eclipsed by the materials from the ablating companions during (some of) the observations. New search observations should be made at the updated position. 

\section{Discussion}
In various aspects from radio through GeV $\gamma$-rays, we have shown that \src\ is well consistent with a redback MSP binary: 

\begin{itemize}\setlength\itemsep{0.0cm}
\item The system exhibited a clear orbital modulation in optical with a compact orbit of $P=9.3$ hours and $K_2=272$~km s$^{-1}$, which is common among the known redback systems. 
\item The X-ray counterpart has a hard spectral index (i.e., $\Gamma_X\approx1.4$) comparable to other redbacks, though the spectral index is uncertain, and deeper X-ray observations are needed. 
\item The $\gamma$-ray flux is mostly stable on a monthly time scale and a significant curvature is found in the LAT spectrum. Both are signature features observed in many LAT-detected pulsars. 
\item It has a pulsar-like radio counterpart with a steep spectral index of $\alpha\approx-1.6$. 
\end{itemize}

These strongly suggest that the $\gamma$-ray source is a new redback MSP binary. We also comment that \src\ is unlikely to be a black widow MSP, since the source needs to be very nearby (i.e., $d=200$ pc) to make the G-type secondary small in size (i.e., $\sim0.1R_\sun$). At this small distance, the X-ray and $\gamma$-ray luminosities would be as low as as $L_X\approx1.5\times10^{30}$\lum\ and $L_\gamma\approx8\times10^{31}$\lum, which are far too low to produce the huge ``day'' and ``night'' temperature difference seen in the Gemini spectra (i.e., $L\sim10^{34}$\lum\ is required; will be discussed in \S\ref{sec:model}). 
In addition, this close distance also contradicts the parallax of \src\ measured by \textit{Gaia} (see \S\ref{sec:bin}). 
In contrast, a redback MSP scenario can provide a self-consistent picture for the multi-wavelength observations. 

\subsection{Basic Properties of the Binary}
\label{sec:bin}
The observed kinematics of the secondary yield the mass function $f(M) = P_{\rm spec} K_2^3/(2 \pi G) = M_1 (\textrm{sin} \, i)^3/(1+q)^2$ for gravitational constant $G$, inclination angle $i$, and mass ratio $q = M_2/M_1$. For SSS~J095527.8$-$394752, we find $f(M) = 0.81\pm0.04 M_{\odot}$. Assuming that the primary is a neutron star with a maximum mass of $2.0 M_{\odot}$, then the observed mass function implies $i > 48^{\circ}$ (for $M_1 = 1.4 M_{\odot}$, $i > 56^{\circ}$).
Given that most redbacks have secondaries with masses of at least $0.2 M_{\odot}$, the most likely range of the inclinations is somewhat more restricted, $i > 52^{\circ}$ ($M_1 = 2.0 M_{\odot}$) to $i > 66^{\circ}$ ($M_1 = 1.4 M_{\odot}$). 
Under these circumstances, the separation between the two binary members can be well restricted to $a\sim (1.8$ -- $1.9)\times 10^{11}$cm (equivalent to 2.6 -- 2.7 $R_\sun$) for $M_2=0.1$ -- $0.4M_\sun$. 

Although not exclusively, normal redback systems often have luminosities of $L_X\lesssim10^{32}$\lum\ (see, e.g., \citealt{2014ApJ...795...72L}), putting a weak constraint of $d\lesssim1.7$~kpc on the distance for the redback MSP candidate. 
Another constraint comes from the parallax information from the second \textit{Gaia} data release (GDR2; \citealt{2016A&A...595A...1G,2018arXiv180409365G}), $\varpi=0.36\pm0.17$~mas (corrected for the global zero point of $-0.029$~mas; \citealt{2018arXiv180409366L})\footnote{In GDR2, all parallaxes are computed by assuming that the sources are single stars. This could cause additional systematic uncertainties for binary systems.}, which can be converted to a geometric distance of $d=2.4^{+1.2}_{-0.7}$~kpc (probability contained: 68\%; \citealt{2018arXiv180410121B,2018arXiv180409376L}). 
Although these constraints are not totally consistent with each other, they both roughly agree $d\approx1.7$~kpc. Assuming $d=1.7$~kpc, the $V_{\rm css}\sim17$~mag G-type companion would have a radius of $R_c\sim 0.7R_{\odot}$\footnote{$V_{\rm css}\approx17$~mag was inferred from the best-fit magnitude at night presented in \S\ref{sec:crts} (without irradiation of the pulsar).  An extinction correction with the best-fit \nh\ of $5.2\times10^{21}$\cm\ obtained from X-ray spectral fitting \citep{2009MNRAS.400.2050G} has been applied. If \nh\ $=10^{21}$\cm\ is assumed, the $V_{\rm css}$ magnitude will increase to 18.6~mag (fainter) with a smaller inferred companion size of $R_c\sim0.3R_{\odot}$.}. This stellar size is indeed in line with the fact that no obvious ellipsoidal variation is seen in the CSS light curve (average error: 0.3~mag), given that only weak ellipsoidal variability can be created by a $R_c\sim 0.7R_{\odot}$ secondary for \src\ (e.g., peak-to-peak amplitude $\sim0.4$~mag for $M_1=1.4M_\sun$ and $M_2=0.4M_\sun$, computed with the ELC code; \citealt{2000A&A...364..265O}). 

\subsection{X-Ray Orbital Modulation?}
\label{sec:xmod}
To look for the X-ray modulation, we made an XRT light curve folded on the orbital period of $P_{\rm joint}=0.3873279$ d (Figure~\ref{fig_fold_lc}). 
As we have mentioned in \S\ref{sec:swift}, the X-ray counterpart was not detected in every \textit{Swift}/XRT observation. 
For better visualization, we plotted the bins with different colors based on whether the source was detected in a single full observation (detected: red; undetected: black). 
For all non-detection data bins, we present Bayesian bins (i.e., net count rates handled by a Bayesian approach, even for bins with zero net count rate; \citealt{1991ApJ...374..344K}) in addition to 95\% upper limits to scratch the possible orbital modulation in X-rays (Figure~\ref{fig_fold_lc}d). 

While the black dataset does not show evidence for a modulation, the red one is likely orbitally modulated with stronger X-ray emission seen in phase 0.6--0.8 (Figure~\ref{fig_fold_lc}d). This tentative X-ray peak seemingly aligns with the optical/UV peak at phase 0.75, which is the superior conjunction of the pulsar binary. Interestingly, X-ray modulation peaks around the superior conjunctions were commonly observed in redbacks, e.g., PSR J2129$-$0429 \citep{2015ApJ...801L..27H} and PSR J1723$-$2837 \citep{2017ApJ...839..130K}. 

The poor data quality could be the reason to explain why the modulation was unseen in the black dataset (see \S\ref{sec:swift}). Alternatively, it may imply a slightly unstable orbital modulation of the binary. In fact, it has been shown in PSR~J1723$-$2837 that the X-ray orbital modulation of redback MSP binaries can change slightly from orbit to orbit due to, e.g., the wind instability of the companion (see Figure 1 in \citealt{2017ApJ...839..130K}). Further X-ray observations would remove the ambiguity. 

\subsubsection{No Sign of Gamma-Ray Orbital Modulation}
We also searched for possible $\gamma$-ray orbital modulation at $>100$ MeV by folding the $\gamma$-ray photons accordingly. Different aperture radii from $0\fdg1$ to $1^\circ$ were tried, but no evidence for an orbital modulation was found. 

\subsection{Interpretation for the High-Energy Emission}
\label{sec:model}
Spin-down power is the major energy source for rotation-powered pulsars. Although the pulsations as well as the spin-down rate of \src\ have not been detected yet, we can still infer the spin-down power by measuring the pulsar irradiation on the companion, if the spin-down power and the irradiation power are approximately the same. With the assumptions that the radiation/pulsar wind from the pulsar are both isotropic and the irradiated hemisphere of the companion is uniformly heated, the spin-down power of \src\ can be estimated from 
\begin{equation}
\epsilon L_{sd}\sim \sigma_B (T_{d}^4-T_{n}^4)\frac{2\pi R_{c}^2}{\delta \Omega},
\label{eq:heat}
\end{equation}
where $\epsilon$ is the heating efficiency, $\sigma_B$ is the Stefan-Boltzmann constant, $T_{d}\sim 8000$K and $T_{n}\sim5700$K are the ``day'' and ``night'' temperatures, respectively, and $\delta \Omega$ is the fraction of the sky covered by the companion star seen from the pulsar, which can be written as $\delta \Omega\sim \pi (R_c/a)^2\sim 0.47$ for $a\sim 1.8\times 10^{11}$cm (corresponding to $M_2\approx0.2_\sun$). 
The inferred spin-down power is $L_{sd}\sim\frac{1}{\epsilon}\times 10^{34}\,{\rm erg~s^{-1}}$, a very typical value for MSPs. 

\begin{figure}
\centering
\includegraphics[width=0.48\textwidth]{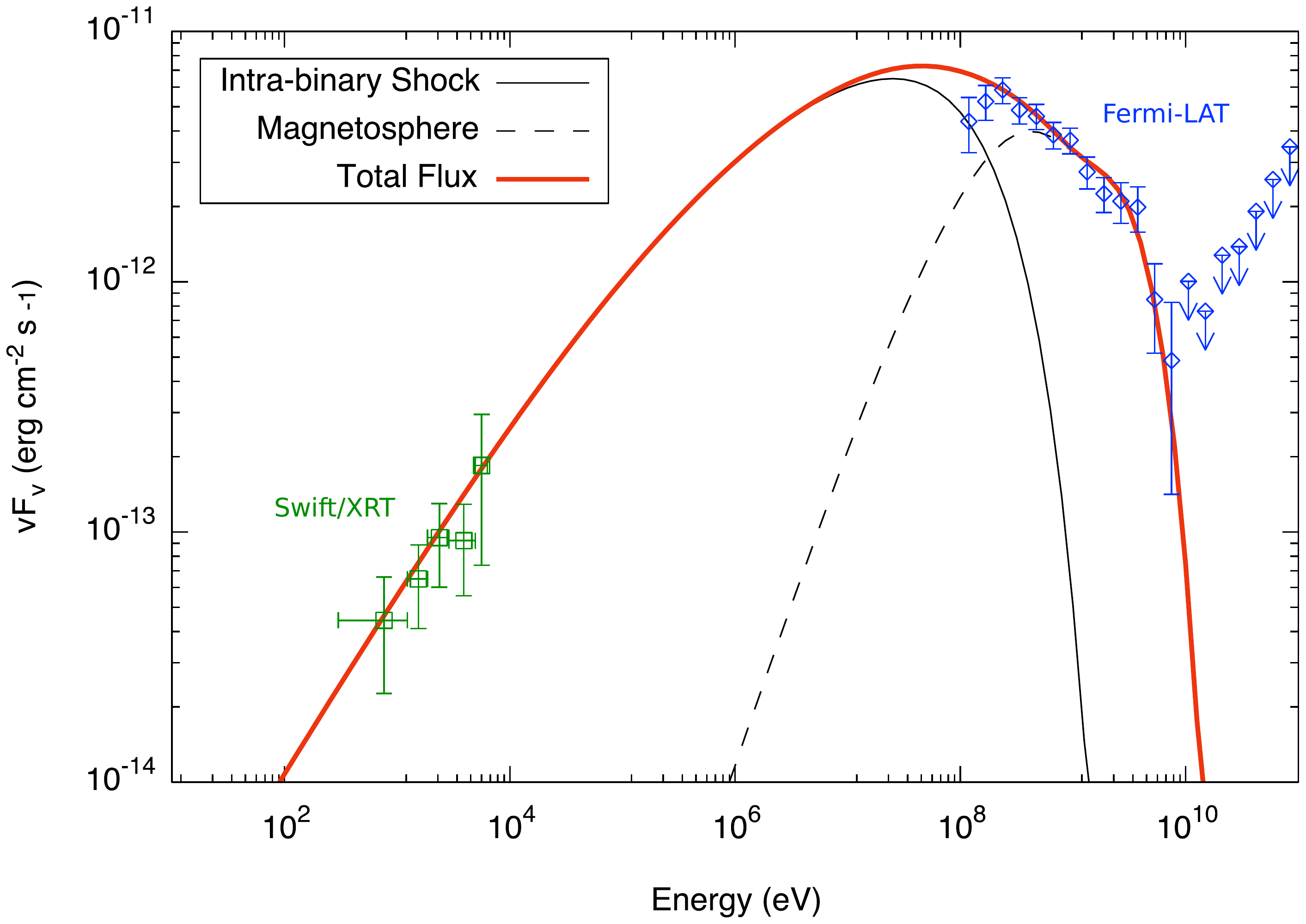}
\caption{Spectral energy distribution of \src\ in the soft X-ray band (\textit{Swift}/XRT; 0.3--10~keV; absorption corrected with \nh\ $=5.2\times10^{21}$\cm) and the GeV $\gamma$-ray band (\textit{Fermi}-LAT; 0.1--100~GeV). The solid and the dashed lines are the energy spectra from the intra-binary shock and the outer gap of the pulsar magnetosphere, respectively. See \S\ref{sec:model} for more details of the model. 
}
\label{fig:sed}
\end{figure} 
 
Presumably, the X-ray and GeV $\gamma$-ray emission of \src\ originates from the intra-binary shock and the pulsar magnetosphere \citep{2014ApJ...785..131T,2014ApJ...797..111L,2017ApJ...850..100A}, if it is a redback MSP, i.e., the shock-accelerated electrons and positrons emit X-ray/soft $\gamma$-rays via synchrotron radiation, and even TeV photons via inverse-Compton scattering off the stellar photons. 

The outer gap emission from the pulsar magnetosphere \citep{1986ApJ...300..500C} will contribute significantly in the LAT energy domain, especially above 1~GeV. We computed the outer gap spectrum using the emission model developed by \cite{2010ApJ...720..178W} with the spin-down power of
\begin{equation}
L_{sd}=3.8\times 10^{35} \Big(\frac{P_s}{1\,{\rm ms}}\Big)^{-4}\Big(\frac{B_s}{10^8\,{\rm G}}\Big)^2\,{\rm erg\,s}^{-1}, 
\end{equation}
which gives the strength of the surface dipole field $B_s\sim1.5\times10^{8}$~G for a spin period of $P_s\sim 3\,$ms and $L_{sd}=2\times10^{34}$\lum\ ($\epsilon\sim0.5$ assumed). 
The magnetospheric emission from the outer gap is highly related to the gap fraction, i.e., $L_{\rm gap,\gamma}\sim f_{\rm gap}^3L_{sd}$, where the gap fraction $f_{\rm gap}$ is defined by the ratio of the gap thickness to the light cylinder radius at the light cylinder. 
For MSPs, the gap fraction is empirically found to be $f_{\rm gap}>0.3$, as the LAT $\gamma$-ray luminosities are often $>10$\% of the spin-down powers \citep{2013ApJS..208...17A}. In the model calculation for \src, we assumed a gap fraction of $f_{\rm gap}\sim 0.5$ to explain the observed luminosity with $d=1.7$~kpc. As long as we fix the spin-down luminosity and the gap fraction, the predicted $\gamma$-ray luminosity is not sensitive to the spin period. 

For the intra-binary shock contribution, we first roughly estimate the ratio of the momenta of the stellar wind and the pulsar wind ($\eta_b$) from the X-ray light curve. 
It has been suggested that the X-ray orbital modulation of redback systems is caused by the Doppler boosting of the post-shocked wind (e.g., PSRs~J1023+0038 and J2129$-$0429; \citealt{2014ApJ...797..111L,doi:10.1093/mnras/sty1459}). If the X-ray modulation of \src\ is genuine, then the shock cone has to be wrapping the pulsar, such that the X-ray emission can be Doppler boosted to create an X-ray flux maximum when the companion is behind the pulsar (phase = 0.75 in Figure~\ref{fig_fold_lc}). In this case, the stellar wind should have a larger momentum than the pulsar wind. 
We therefore assumed $\eta_b=7$, which was also used to explain the X/$\gamma$-ray properties of the redback MSP~J1023+0038 \citep{2014ApJ...797..111L}. 
The shock geometry with $\eta_b=7$ is calculated with the method discussed by \cite{1996ApJ...469..729C}, with the magnetic field strength at the shock parametrized by the magnetization $\sigma$. Besides the shock geometry, this magnetization parameter also controls the X/$\gamma$-ray luminosities from the intra-binary shock once the spin-down power is fixed at a certain value (i.e., $L_{sd}=2\times10^{34}$\lum), as the synchrotron luminosity is proportional to the square of the magnetic field strength, which depends on the magnetization as $\sigma^{1/2}$. For \src, we applied $\sigma=0.1$ to match the observed X-ray luminosity. 

The initial energies of the accelerated particles are assumed to be power-law distributed. The maximum energy of the particles is determined by balancing the synchrotron loss time scale and the acceleration time scale, yielding the Lorentz factor of $\gamma_{max}=[9m_e^2c^4/(4e^3B)]^{1/2}$ at maximum, while $\gamma_{min}=10^4$ is assumed for the minimum Lorentz factor. We finally solve the evolution of the energies of the particles under synchrotron radiation (inverse-Compton scattering loss is negligible for the MSP binaries) and calculate the correspinding intra-binary shock emission. Details of the calculation can be found in \cite{2014ApJ...785..131T} and \cite{2014ApJ...797..111L}. 

Figure \ref{fig:sed} shows the calculated emission components from the intra-binary shock and the outer gap, which match the \textit{Swift}/XRT and \textit{Fermi}-LAT spectra very well. Despite the ideal consistency, we emphasize that the parameter space for this system has not been fully explored, and therefore this specific parameter set is just one example that appears to fit the data. Additional measurements for the key parameters, such as $L_{sd}$ and $P_s$, are essentially required to accurately capture the full physical behavior of the system in the future. 

Note that \cite{2015ApJ...810...85C} classified \src\ as a ``poorer pulsar candidate'' because of its ``monotonically decreasing'' (with energies) LAT spectrum, which is theoretically unfavoured for a pulsar. With our intra-binary shock model, we have demonstrated that, even though the magnetospheric emission of a redback pulsar (the outer gap model for instance) peaks at $\sim1$~GeV (in the frame of $\nu F_\nu$), the spectrum can be contaminated at low energies by the $\sim0.1$~GeV shock emission, making the $\gamma$-ray counterpart non-pulsar-like. 
Apparently, $\gamma$-ray spectra are not good as a pulsar indicator alone when very strong intra-binary shock emission is observed.

\subsection{The Largely-Shifted LAT Localization}
We should not expect 100\% accuracy for 95\% error ellipses by definition, however, the case of \src\ is quite extreme---the 3FGL and the updated error regions are separated by 6\arcmin\ and just barely touch each other by the edges (Figure \ref{fig:dss})---which may worth a brief discussion. 
We compared the 3FGL localization with the previous results in 1/2FGL, the 1FGL 95\% error ellipse is big enough to encompass all the localizations, including the new best-fit position, while the ones in 2FGL (which has the X-ray/optical/radio counterpart inside) and 3FGL are more consistent with each other. 
To ensure the reliability of the new LAT localization, we visually checked the count map and the TS map, which both confirm the new best-fit position. 
Furthermore, our new error circle is almost identical to that in FL8Y (i.e., FL8Y~J0955.4$-$3949; Figure \ref{fig:dss}). 

 
Obviously, the great improvement is given by a much higher quality of data as more than 9 years of Pass 8 data were used. 
At the same time, \src\ is a relatively bright and significant $\gamma$-ray source in 3FGL with a detection significance of $19\sigma$, reflecting that the unsatisfactory localization is unlikely caused by insufficient quality of the data used in the catalog. Strong variability of nearby sources can possibly affect the best-fit localization of an object, however, no cataloged variable $\gamma$-ray source can be found within 4$^\circ$ from \src. Alternatively, the Galactic diffuse emission in this area may not be well modelled in 3FGL, leading to residual emission that could cause the offset. Though \src\ is about $11^\circ$ away from the Galactic plane, there is still plenty of Galactic structure to make this scenario possible. 
Lastly, the relatively soft $\gamma$-ray spectrum of \src\ (the photon index is 2.54 in 3FGL; \citealt{2015ApJS..218...23A}) could also account for the underestimated positional uncertainty. 

In any case, our \textit{Fermi}-LAT study on \src\ clearly shows that the 95\% error ellipses from the likelihood analysis can be underestimated sometimes, even for bright sources with significant detections. While we believe that most of the 3FGL error ellipses are still reliable, expanding the regions of interest slightly beyond the LAT error ellipses could be a wise strategy when identifying unassociated \textit{Fermi}-LAT sources with multi-wavelength observations.

\begin{acknowledgements}
The \textit{Fermi} LAT Collaboration acknowledges generous ongoing support from a number of agencies and institutes that have supported both the development and the operation of the LAT, as well as scientific data analysis. These include the National Aeronautics and Space Administration and the Department of Energy in the United States; the Commissariat \`a l'Energie Atomiqueand and the Centre National de la Recherche Scientifique/Institut National de Physique Nucl\'eaire et de Physique des Particules in France; the Agenzia Spaziale Italiana and the Istituto Nazionale di Fisica Nucleare in Italy; the Ministry of Education, Culture, Sports, Science and Technology (MEXT), High Energy Accelerator Research Organization (KEK), and Japan Aerospace Exploration Agency (JAXA) in Japan; and the K.~A.~Wallenberg Foundation, the Swedish Research Council, and the Swedish National Space Board in Sweden. Additional support for science analysis during the operations phase is gratefully acknowledged from the Istituto Nazionale di Astrofisica in Italy and the Centre National d'\'Etudes Spatiales in France.

X.H. is supported by the National Natural Science Foundation of China through grants 11503078 and 11661161010.
J.S. acknowledges support from a Packard Fellowship. Support from NSF grant AST-1714825 and NASA grant 80NSSC17K0507 is gratefully acknowledged.
J.T. is supported by the National Science Foundation of China (NSFC) under 11573010, U1631103 and 11661161010.
A.K.H.K. is supported by the Ministry of Science and Technology of the Republic of China (Taiwan) through grant 105-2119-M-007-028-MY3.
C.Y.H. is supported by the National Research Foundation of Korea grant 2016R1A5A1013277.
K.S.C. is supported by GRF grant under 17302315.

Support for this work was partially provided by the National Aeronautics and Space Administration through Chandra Award Number GO7-18036X issued by the Chandra X-ray Observatory Center, which is operated by the Smithsonian Astrophysical Observatory for and on behalf of the National Aeronautics Space Administration under contract NAS8-03060. 
This work made use of data supplied by the UK Swift Science Data Centre at the University of Leicester.
This work is also based on observations obtained at the Gemini Observatory (Program ID: GS-2017B-FT-15), which is operated by the Association of Universities for Research in Astronomy, Inc., under a cooperative agreement with the NSF on behalf of the Gemini partnership: the National Science Foundation (United States), the National Research Council (Canada), CONICYT (Chile), Ministerio de Ciencia, Tecnolog\'{i}a e Innovaci\'{o}n Productiva (Argentina), and Minist\'{e}rio da Ci\^{e}ncia, Tecnologia e Inova\c{c}\~{a}o (Brazil).
The CSS survey is funded by the National Aeronautics and Space
Administration under Grant No. NNG05GF22G issued through the Science
Mission Directorate Near-Earth Objects Observations Program. The CRTS
survey is supported by the U.S.~National Science Foundation under
grants AST-0909182 and AST-1313422. 
This work has made use of data from the European Space Agency (ESA) mission
{\it Gaia} (\url{https://www.cosmos.esa.int/gaia}), processed by the {\it Gaia}
Data Processing and Analysis Consortium (DPAC,
\url{https://www.cosmos.esa.int/web/gaia/dpac/consortium}). Funding for the DPAC
has been provided by national institutions, in particular the institutions
participating in the {\it Gaia} Multilateral Agreement.
\end{acknowledgements}
\textit{Facilities}: \facility{Fermi, Swift, SOAR, Gemini:South}

\bibliography{j0954}
\end{document}